\documentclass[conference]{IEEEtran}
\IEEEoverridecommandlockouts
\pdfoutput=1
\usepackage{cite}
\usepackage{amsmath,amssymb,amsfonts}
\usepackage{algorithmic}
\usepackage{graphicx}
\usepackage{textcomp}
\usepackage{xcolor}

\usepackage{amsmath}
\usepackage{graphicx}
\usepackage[ruled, linesnumbered]{algorithm2e}
\begin{document}

\newtheorem{definition}{Definition}[section]
\title{Discrete Search in Heterogeneous Integer Spaces for Automated Choice of Parameters using Correct-by-Construction Methods
}

\author{\IEEEauthorblockN{Omar Radwan}
\IEEEauthorblockA{oradwan@usc.edu \\
oradwan@alumni.usc.edu\\
\textit{Viterbi School of Engineering} \\
\textit{University of Southern California}}
\and
\IEEEauthorblockN{Yilin Zhang}
\IEEEauthorblockA{yilinz80@usc.edu\\
\textit{Viterbi School of Engineering} \\
\textit{University of Southern California}}
\and
\IEEEauthorblockN{Luca Geretti}
\IEEEauthorblockA{geretti@usc.edu\\
\textit{Viterbi School of Engineering} \\
\textit{University of Southern California}}

}

\maketitle

\begin{abstract}
Discrete Search of integer spaces for tool parameter values provides a powerful methodology for modeling and finding a heuristically optimal parameter list for a given system. Current tools and implementations that exist focus primarily on homogeneous tool parameters, and the implementations for heterogeneous tool parameters is lacking. In this paper we introduce a correct-by-construction method of heterogeneous parameter reachability and validity search, and further outline the implementation as well as a demonstration using examples of heterogeneous systems that this tool can be used for. 
\end{abstract}
\pagestyle{plain}

\section{Introduction}

\subsection{Premise}
Discrete Search of integer spaces provides a powerful mechanism through which to explore the reachable set of a given system. Current design cools work primarily for homogeneous parameter spaces, and mapping a heterogeneous  parameter space into the integer domain would provide a strong backbone for both performance and allow for a wide range of uses in many hybrid systems as well as hybrid parameters that are contained within a single system. There are precautions that would need to be taken for hybrid systems, which primarily consist of having unsafe states, that even though they are reachable, they would be considered to be unsafe in a real-world implementation, as well dependencies between variables, that could transcend the homogeneous dependencies that are trivial (i.e. comparing two integers together as compared to a comparison between floating point and Boolean). There also would exist optimal state locations of the parameter set, and those would be modeled using an arbitrary cost function. 

\subsection{Related Work}
Related work consists primarily of homogeneous tool parameter exploration implementations, and those concern themselves primarily with arriving at the reachable set primarily for homogeneous parameter sets. This would include the tool Ariadne \cite{Luca2022}, which has features built-in that allow it to find an approximation of the given reachable set by giving by controlling the growth of the approximation error. 

One other concern that arises when attempting to model heterogeneous parameters in integer spaces is the problem of solvability within bounded time with close approximation, and as outlined in \cite{Conforti2014}, there does exist a finite bound for finite discovery. There was a foray into unbounded analysis, but that is infeasible given the constraints and would be too computationally exhaustive. Another issue that comes up is discrete versus non-discrete evolution in terms of time, and this was a problem resolved by setting as a condition that there can only exist discrete time steps and discrete evolution. 

\begin{figure}[!htbp]
\includegraphics[width=8cm]{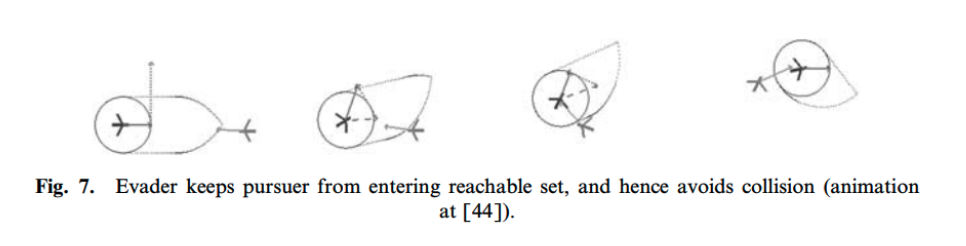}
\caption{From Citation \cite{mitchell2003overapproximating}}
\end{figure}

\subsection{Our Approach}

For the implementation demonstrated in this paper, we focus on a number of contributions that create a fast and efficient method of finding the optimal set given existing constraints and cost. We also define a semantic format that supports the representation of heterogeneous parameters, which better suits it for discrete search along hybrid domains. 

For exploring the adjacent set space from our beginning iteration point(initial state), there are a number of possible implementation decisions that would need to be made on how best to explore the reachable set given the constraints. The path that we decided on was to create a correct by construction approach, that would allow the exploration tool to only explore the reachable set that is also valid given the constraints and dependencies that are supplied. Our flow is as follows: given a parameter list which can consist of integer, Boolean, and composite parameters, as well as a list of constraints and dependencies between variables, and a cost function, we aim to find a valid parameter state that satisfies all of our given requirements. 

For our implementation, we split our computational engine into two general algorithms. Our first algorithm involves computing a correct-by-construction interval for a given parameter given our requirements, and our current state when it comes to other parameters that exist within our set space. The second algorithm is our step-by-step evolution iteration across the set space of the parameter list based on the computation of local optimal cost. 

Compared to existing and related works, our approach has the following contributions:
\begin{itemize}
\item Developed a representation for heterogeneous parameter sets that allows for the discretization of all parameters and results in the ability for integer space exploration for all relevant types
\item Created a correct-by-construction approach to not only finding the reachable set of a given parameter set, but also allowing the inclusion of heterogeneous inter-parameter dependencies and assertions. 
\item Designed a method of evolution that allows for quick computation of adjacent states for a given set of already locally-optimal parameter instances with a method of back-tracing and reset if arriving at an invalid location
\item Demonstrate the applicability and the versatility of our implementation on two examples that involve computing minimum cost for a computer architecture design and a re-programmable logic circuit with a demonstration of the implementation of pseudo-Boolean constraints

\end{itemize}

\section{Implementation}
\subsection{Environment and Language Considerations}

We decided on implementing our design in Python\cite{python}, the reason for that being that Python allows a host of libraries and type-interfacing that would allow us to quickly prototype, verify, and extend during testing. We also chose Python for the reason of being able to interface easily with JSON\cite{JSON}, which is our input-format of choice. JSON was chosen due to its status as being very well-adopted and would provide an easy interface for other CAD tools to create tool-parameter sets for analysis using our program. 

We also use a number of Python libraries to do the necessary computations that are required for our implementation. A special recognition is deserved of Numpy\cite{numpy}, which is a library that allows for very quick computation of intervals, arrays, and sets. Since we are operating in the integer domain, integer arrays using Numpy libraries make the cost of computation a significantly smaller area of concern during implementation. 

\subsection{Motivation for Design}
To better improve the performance of discrete search in heterogeneous space, there do exist a number of limitations. Firstly, a slight weakness exists in parsing string type assertions and evaluating them in a computationally static format as opposed to extensive abstract syntax trees and symbolic interval computation. Secondly, considering various typed parameters and assertion relations, it is necessary to have a uniform interface design such that algorithm implementation is isolated with complicated typed transformation, which is why JSON was selected, which could become unwieldy if enumerated or vector parameters which to be considered. In this case, a tool that would generate a statically-enumerated JSON format that is acceptable to our program would be required. 

\subsection{Evolution Algorithm}
In this section, we introduce how the program will explore feasible set constrained by assertions. The JSON format input will be interpreted and loaded into our program. For the sake of generality, we assume that there are $n$ parameters denoted as $x_1, \cdots x_n$. First of all, for each parameter $x_i$, we randomly generate $N-1$ valid neighboring points. For the random sampling of these points, we experimented with a couple methods. One was uniform sampling from the valid interval, the other two where linear and square weighted sampling with respect of distance from the interval. After these were tested, we found that square weighting was the most effective, and we will demonstrate these findings during our examples. With $x_i$ itself, these $N$ points form a list $\{x_i^j\}_{j=1}^N$. In total there are $n$ lists.

During the evolution process, each point will randomly generate a neighboring point from its valid set. Therefore, all $n\cdot N$ points will generate another $n$ new lists. Without loss of generality, we denote these $n$ new lists as $\{x_i^j\}_{j=N+1}^{2N}$. Next we  the original list and new list with the same footnote $i$ to get $n$ new list $\{x_i^j\}_{j=1}^{2N}$. From these $n$ list, we evaluate $2N$ cost function values as $\{c_j = F(x_1^j, x_2^j, \cdots x_n^j) \mid j = 1, \cdots, 2N\}$. For these $2N$ cost values, we keep the smaller half and corresponding parameter values to form $n$ new lists. Repeat the above steps until the ending requirements are satisfied. A pseudo-code for this algorithm can be found at Algorithm \ref{algorithm1}

\subsection{Approach for feasibility checking between heterogeneous parameters}

For defining the the set of parameters that would exist for a given system, we supply two atomic types and one composite type:
\begin{enumerate}

\item Integer type 

\item Boolean type 

\item Composite type

\end{enumerate}

Integers exist in the Integer domain, and Boolean's likewise in the Boolean domain. Composites are different in that they are modeled like an array, given a composite parameter \(C\), \(C\) can contain any number of composites, Boolean's, and integers. This allows the modeling of parameters that cannot be modeled as strictly scalar integer or Boolean values. Floats, complex numbers, and vectors are all examples of what can be modeled as a composite set. Furthermore, to maintain the desired behavior of these parameters, the constraint paradigm that we introduce allows us to describe the behavior of how these composite parameters undergo evolution. 

As an example, take \(Cube\), which of type composite, and it is defined by 3-equal length sides \(x,y,z\), such that \(Cube(t)= \{x,y,z \in \mathbb{Z}, x==y==z \} \forall t \) where \( t \) is time-step during evolution. For the case of this parameter, the instantiation of the of the domain of each sub-parameter would go with the parameter declarations, while the instantiation of the constraint that is intrinsic to cubes would be added to the constraints field that is given. 

This paradigm of allowing composite parameters to have unique behaviors could lead to invalid states during evolution, if one sub-parameter undergoes evolution independently and is now not equal to the other two, that would lead to an undesirable state. For this reason correct-by-construction interval generation for each of the sub-parameters is done with all assertions and constraints in mind. 

One note on using composite parameters to model floating point numbers. Initially during development we had planned to incorporate a floating point type, however the tediousness of setting properties for floating point as an atomic type is redundant as all the properties of a floating point value(mantissa, exponent, significant figures) can be modeled as sub-parameters of a composite value, and the user can specify the desired constraints and behaviors for comparison and incorporation between the composite-ized floating point value and other parameters. 

\subsection{Feasibility Checking given Constraints}
In this section, a detailed explanation about how to construct valid neighboring set is given. Suppose that there are $m$ assertions $\{\mathcal{A}_i\}_{i=1}^m$ on $n$ parameters. For each parameter $x_i$, assertions containing $x_i$ are selected out of $m$, which is $\{\mathcal{A}_k \mid x_i \in \mathcal{A}_k \}$. Next, iterate through other parameters and apply their values into these assertions. Finally, Intersect all the intervals after evaluating the assertions to get the final interval. A new value for $x_i$ is sampled randomly from the final interval based on the square of their distance to $x_i$. By default, values closer to $x_i$ have higher probabilities to be selected. More details can be found in Algorithm \ref{algorithm2}

\begin{algorithm}
    \caption{Evolution of Adjacent Optimal Cost}
    \label{algorithm1}
    \KwIn{List of variables $L_v$, Iterating parameter $T$, List of assertions $L_a$, Cost function $F$}
    \KwOut{Optimal value of variables $L_v^*$}
    \BlankLine

    //This is for initial value selection, since we need to enter the set space is what we presume to be a valid point
    \ForEach{\textnormal{$v$ in $L_v$}} {
        $v$ := Sample Uniform Distribution(Lower Bound of $v$, Upper Bound of $v$)
        
        Construct $V_i$ as the set of $n$ sample of $v_i$
    }   
    \While{\textnormal{$T <= K$}}{
        \ForEach{\textnormal{variable $v_i$ in $L_v$}}{
            $V_i$ is the set of $n$ values of $v_i$
            $S_{v_i}=$get\_intersect\_of\_all\_valid\_intervals$(L_a, L_v, v_i)$.
          
            $S_{v_i sorted}$ = Arrange by incrementing closeness to value of $a_k$
            
            $Weights_{S_{v_i sorted}}$ = array from 0 to length of $S_{v_i sorted}$ 
            
            \ForEach{ $w$ in $Weights_{S_{v_i sorted}}$}{
            
            $w$ = (length of $S_{v_i sorted}$ - index of $w$)$^2$
            
            }   
            Use weighted sampling of $Weights_{S_{v_i sorted}}$  to randomly sample $n$ new values of $v_i$ from $S_{v_i sorted}$.

            Append these $n$ values into $V_i$
            
            Construct $n$ new list of variables $\{L_v^j\}_{j=1}^n$, $L_v^j[i] = L_v[i]$.

            Pick $L_v^k$ with minimum $F(L_v^k)$ in $\{L_v^j\}_{i=j}^n$.

            Update $L_v[i] = L_v^k[i]$.

            Delete $v_i$ in $V_i$ with $n$ highest cost values.

            Update $T$.
        }
    }  
    \Return $L_v$
\end{algorithm}

\begin{algorithm}
    \caption{Get Intersect of All Valid Points in Bounds and Assertions}
    \label{algorithm2}
    \KwIn{List of assertions $L_a$, List of variables $L_v$, Target variable $v_i$}
    \KwOut{All valid set $S_{v_i}$ of $v_i$}
    \BlankLine
    Initialize list of intervals $L_i = [ ]$

    \ForEach{\textnormal{$a_k$ in $L_a$}}{
        \eIf{\textnormal{$v_i$ appears in $a_k$}}{
            \ForEach{\textnormal{$v_j$ in $L_v$}}{
                \eIf{$v_j \neq v_i$}{
                    Plug in value of $v_j$ in $a_k$.
                }{
                    continue
                }
            }
            Append $a_k$ into $L_i$
        }{
            continue
        }
    }


    Transform the intersection of $L_i$ into valid set $S_{v_i}$
    
    \Return $S_{v_i}$

\end{algorithm}

\subsection{Desired Implementation Aspects that Proved Infeasible}
One initial idea that was considered well thought out and feasible was the incorporation of symbolic computation for our constraint and dependency valid interval generation. The Sympy\cite{sympy} library in Python was going to be utilized for this purpose. Though the algorithm was functional, the symbolic computation cost was extremely prohibitive, and was not feasible for a general-use case. After doing much research to attempt to make it feasible, we discovered that even Sympy as an organization recognizes that the substitution and evaluation is cost-prohibitive, and recommends other avenues for repetitive computation. For this reason we had to re-calibrate and find another solution. This solution was to do string replacement of our given parameters with their values into the string representation of our constraints, dependencies, and costs. Then these string representations would be converted into lambda functions that would be operated on by the Numpy array operations. Since Numpy on the back-end uses C libraries to do computation, this lessened our computation time by an order of magnitude, for mostly the same functionality. 

The functionality that is missing is due to the inherent behavioral properties of lambda functions. Symbolic computation was desired as it allowed the incorporation of very rigorous Boolean SAT exploration, but this is not a feature that is possible with the lambda paradigm. Therefore, to allow the extend-ability of Boolean values, fuzzy pseudo-Boolean logic\cite{fuzzylogic} is implemented, which does allow for an adequate semantic representation of Boolean logic. 

\section{Examples of Application}
For an example foray to explore what our program would be able to handle, we decided on two different, yet related, domains.
\subsection{FPGA Synthesis}

\begin{figure}[!htbp]
\includegraphics[scale=0.3]{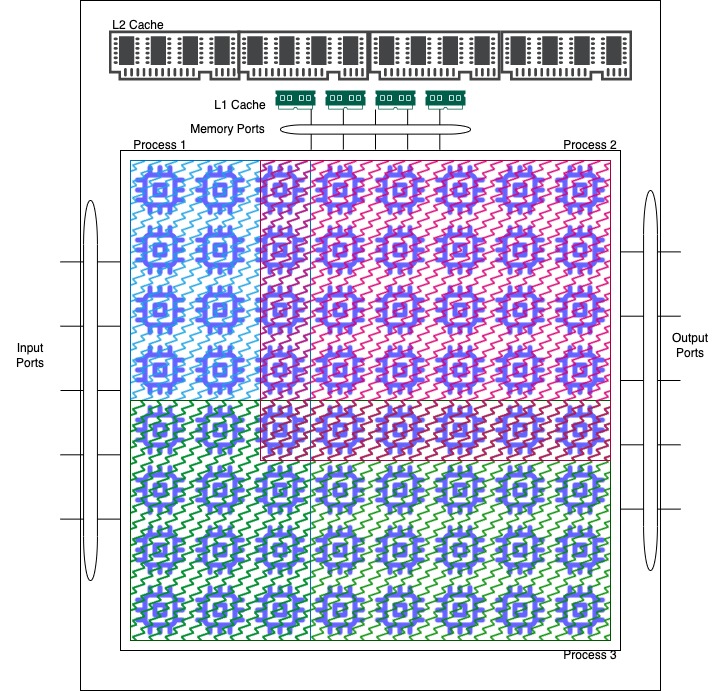}
\caption{Illustration of FPGA Paradigm for Testing Our Implementation}
\label{FPGA_Diagram}
\end{figure}

For our first example(outlined in \ref{FPGA_Diagram}), we decided on modeling our problem as an FPGA cost problem. Given a number of constraints on an FPGA, i.e. memory size, available memory ports, available input and output ports we have \(Routine_{1,2,3}\), and only two of the previously mentioned three can be installed on the FPGA fabric, and depending on which two are loaded onto the fabric, we then must enable a minimum number of memory, I/O, and interconnection ports, as well as have different memory properties. We then created a polynomial cost function of these constraints, in an aim of it becoming nonlinear and make the algorithm demonstrate its effectiveness in traversing the set space while attempting to find the given most optimal cost. 

\begin{figure}[!htbp]
\includegraphics[scale=0.3]{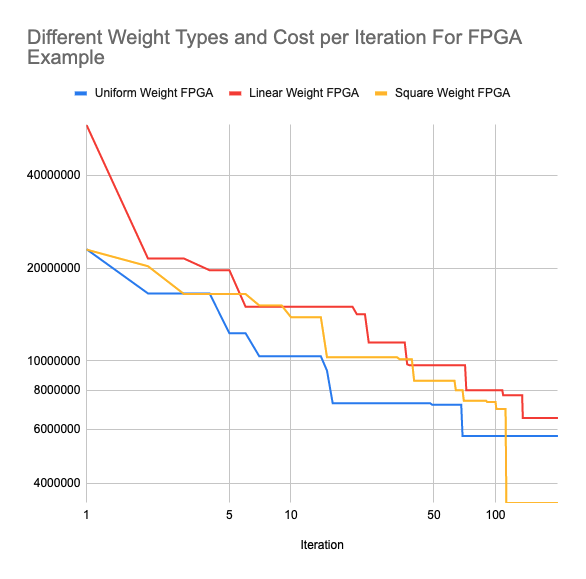}
\caption{Table of the Impact of Different Weights and Effect on Set Exploration for FPGA Example}
\label{FPGA_Performance}
\end{figure}

One highlight of this example is the inclusion of pseudo-Boolean constraints, which manifest in the requirement that only two of the three routines can function at any time, which in terms of cost, creates a piece-wise function. The parameter variation that is generated during random sampling is able to traverse this piece wise function, because even though we generate points using a correct-by-construction approach, in some cases there is no valid interval, and in that case we reset for that specific parameter back to the largest valid interval, and randomly sample that. This allows the program to exit any possible rut that it enters while making an early decision on which Routine set to choose, and so it can backtrack as necessary and choose another Routine set if the specific parameter space undergoing evolution is no longer valid. The results for these are demonstrated in Figure \ref{FPGA_Performance} with different weights for random sampling methods from the valid intervals generated. 

\subsection{Computer Architecture Design }
\begin{figure}[!htbp]
\includegraphics[scale=0.3]{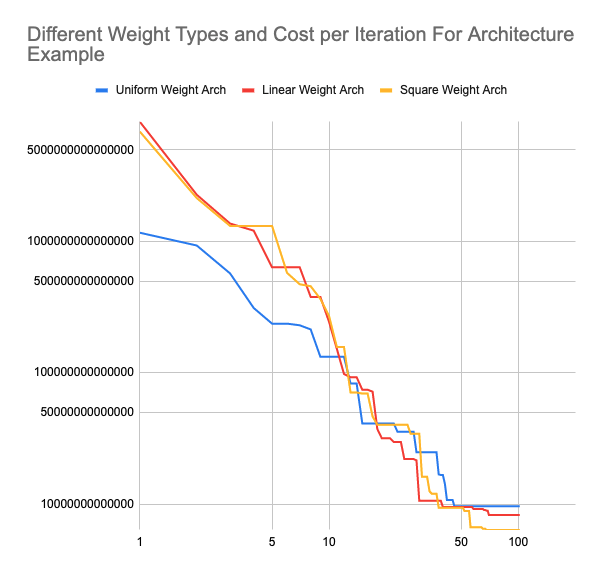}
\caption{Table of the Impact of Different Weights and Effect on Set Exploration for Architecture Example}
\label{Arch_Performance}
\end{figure}

Another example that we used is the creation of of a computer architecture system. During the creation of a new computer architecture, or the generation of a new implementation of an architecture, multiple design decisions must be made with respect to area, inter-connectivity, interface requirements, and transistor count. In this example, we model a simple multi-fetch, multi-execution, processor design. We drafted the requirements in terms of dependencies and constraints, and given the constraints and requirements for the interfaces and inter-connectivity between components, we aim to find the minimal transistor count. This was a more rudimentary design, and it aimed to find the computation limit of our implementation. One thing that we attempted to model was having very large integer sets, and exploring those. Emulating design space exploration for computer architectures with such large intervals was the reason we had to refactor our computation engine from purely symbolic to the lambda paradigm, as the symbolic computation was not able to run search space exploration and computation in a reasonable amount of time with this example. The results for those example are posted in Figure \ref{Arch_Performance}, along with the variation between random sampling methods from the valid intervals generated.

\subsection{Performance and Efficacy}
As aforementioned during the discussion on the implementation, performance was a major bottleneck in our implementation, and there were a number of features that needed to be added to be able to guarantee reasonable performance. The first was the use of lambdas to calculate the valid interval set. The second, which is outlined in the algorithm, is keeping a short list of the least-cost neighbors that exist, and generating new random neighbors from that list. This allows us to have multiple different forays into the search space, and we could possibly arrive to many local minima's, but we only choose the most optimal local minima. Computation time is static across iterations, and there are parameter options to increase or decrease the exhaustiveness of the search depending on the intended use cases. 

We also wanted to verify the efficacy of our design and do the best possible effort into generating the most optimal point. To verify that our results where sane, we ran multiple different instances of both the FPGA and Computer Architecture description JSON files, and averaged those results out, and did this for three different weights for random sampling(uniform, linear weighted, square weighted), and what we found that in all cases, our results for all runs where fairly similar, but there are some noticeable differences worth discussion. 

Firstly, the uniform random search has better performance for lower iterations, and this is because during early stages of evolution, a majority portion of the set space has yet to be explored, and uniform sampling allows us to traverse the majority of the set space early. However after a lot of iterations, the square weighted random sampling from the interval eventually makes us arrive to a more optimal cost, and this is because as more and more of the set space is invalidated, the parameters that are undergoing evolution get much closer to the local optima, and square weighting allows us to more likely sample these local optima and arrive at them at a quicker rate than both uniform and linear random sampling. 

\section{Summary}
To reiterate the major points that have been mentioned throughout this paper, we have created a tool that performs discrete search of integer spaces of mapped heterogeneous parameters to the integer domain, and we utilized correct-by-construction methods to ensure that given constraints and dependencies are met, while attempting to find the most optimal cost. This differs from the previous literature in that it is able to accommodate for heterogeneous data structures and is able to model hybrid systems, while comparatively the existing literature exists primarily for reachability and homogeneous parameter exploration. The main takeaways from this endeavor include that there is a significant divide between the tools that are used in industry, and the potential for tools that could be used to better-optimize processes and methods that are used. The main hurdle for widespread adoption of these methods includes a difficulty of understanding and use, as well as a computational cost-barrier that is evident in very complex systems. 

\subsection{Wish-list of additional features}
One feature that would have been useful to incorporate would have been incorporating a Boolean SAT or SMT solver\cite{SMT}, which would have allowed us to bypass pseudo-Boolean constraints entirely, which are generated heuristically, and instead rigorously solve Boolean equations for all possible solutions. Incorporation a Boolean SAT solver such as Z3 would've been time-prohibitive, but would've allowed for a greater range of expressively for constraints. 
\subsection{Application Files}
Due to space reasons, we do not go into detail on the specifics of the Computer Architecture Example and the FPGA Example. Please contact the authors for more information. 





\section{Some thoughts on optimization and use cases}
Optimization aims at searching for values of $\mathbf{x}$ which minimizes the objective function $f$ bounded by constraints. A general formula of optimization problem is in equation (\ref{optimization}).
\begin{equation}
\begin{aligned}
    &\underset{x}{\arg} \min f(\mathbf{x})
    \\
   s.t.  &\ Constraints\  on\  \mathbf{x}
\end{aligned}
\label{optimization}
\end{equation}
In addition to existing gradient based methods which requires the objective function to be differentiable or even more smooth, discrete search algorithm proposed in this paper achieves a high degree of performance on all kinds of objective functions.

One of the most important features of cyber-physical systems is that they contains both continuous system components and discrete system components. In this case, the constraints may include discrete forms like SATs, and continuous forms like inequalities. Our discrete search algorithm can be used to choose optimal parameters for a cyber-physical system.

\section{Further possible work}
We would like to explore more about the background of reachability analysis. Where does this problem rise from. Moreover, as for existing optimization algorithms like heuristic algorithms, gradient based methods and interior point methods, what are the bottlenecks on applying these algorithms on hybrid system reachability analysis. 

Another topic is the connection between reachability analysis and optimization algorithm. If the reachability problem can be formulated into an optimization problem, then it will be easier to understand the problem from the mathematical properties of objective function.

\bibliography{citations.bib}{}

\begin{thebibliography}{1}

\bibitem{Luca2022}
Luca Geretti, Pieter Collins, Davide Bresolin, and Tiziano Villa.
\newblock Automating numerical parameters along the evolution of a nonlinear
  system.
\newblock In {\em Runtime Verification: 22nd International Conference, RV 2022,
  Tbilisi, Georgia, September 28–30, 2022, Proceedings}, page 336–345,
  Berlin, Heidelberg, 2022. Springer-Verlag.

\bibitem{Conforti2014}
Michele Conforti, Gerard Cornuejols, and Giacomo Zambelli.
\newblock {\em Integer Programming / Michele Conforti, G{\'e}rard
  Cornu{\'e}jols, Giacomo Zambelli}.
\newblock Springer, Cham, 2014.

\bibitem{mitchell2003overapproximating}
Ian~M. Mitchell and Claire~J. Tomlin.
\newblock Overapproximating reachable sets by hamilton-jacobi projections.
\newblock {\em Journal of Scientific Computing}, 19(1):323--346, 2003.

\bibitem{python}
Guido Van~Rossum and Fred~L Drake~Jr.
\newblock {\em Python reference manual}.
\newblock Centrum voor Wiskunde en Informatica Amsterdam, 1995.

\bibitem{JSON}
Felipe Pezoa, Juan~L Reutter, Fernando Suarez, Mart{\'\i}n Ugarte, and Domagoj
  Vrgo{\v{c}}.
\newblock Foundations of json schema.
\newblock In {\em Proceedings of the 25th International Conference on World
  Wide Web}, pages 263--273. International World Wide Web Conferences Steering
  Committee, 2016.

\bibitem{numpy}
Charles~R. Harris, K.~Jarrod Millman, St{\'{e}}fan~J. van~der Walt, Ralf
  Gommers, Pauli Virtanen, David Cournapeau, Eric Wieser, Julian Taylor,
  Sebastian Berg, Nathaniel~J. Smith, Robert Kern, Matti Picus, Stephan Hoyer,
  Marten~H. van Kerkwijk, Matthew Brett, Allan Haldane, Jaime~Fern{\'{a}}ndez
  del R{\'{i}}o, Mark Wiebe, Pearu Peterson, Pierre G{\'{e}}rard-Marchant,
  Kevin Sheppard, Tyler Reddy, Warren Weckesser, Hameer Abbasi, Christoph
  Gohlke, and Travis~E. Oliphant.
\newblock Array programming with {NumPy}.
\newblock {\em Nature}, 585(7825):357--362, September 2020.

\bibitem{sympy}
Sympy Foundation.

\bibitem{fuzzylogic}
Y.~Dote.
\newblock Introduction to fuzzy logic.
\newblock In {\em Proceedings of IECON '95 - 21st Annual Conference on IEEE
  Industrial Electronics}, volume~1, pages 50--56 vol.1, 1995.

\bibitem{SMT}
Leonardo De~Moura and Nikolaj Bj\o{}rner.
\newblock Satisfiability modulo theories: Introduction and applications.
\newblock {\em Commun. ACM}, 54(9):69–77, sep 2011.

\end{thebibliography}
\bibliographystyle{unsrt}

\end{document}